\documentclass{emulateapj}

\usepackage{natbib}

\slugcomment{To Appear in ApJ Letters}

\shorttitle{The SN 2011dh/PTF11eon Progenitor}
\shortauthors{Van Dyk et al.}

\begin{document}

\title{The Progenitor of Supernova 2011dh/PTF11eon in Messier 51}

\author{Schuyler D.~Van Dyk\altaffilmark{1},
 Weidong Li\altaffilmark{2},
  S.~Bradley Cenko\altaffilmark{2},
 Mansi M.~Kasliwal\altaffilmark{3},
 Assaf Horesh\altaffilmark{3}, 
 Eran O.~Ofek\altaffilmark{3,4},
 Adam L. Kraus\altaffilmark{5,6}, 
 Jeffrey M.~Silverman\altaffilmark{2},
 Iair Arcavi\altaffilmark{7},
 Alexei V.~Filippenko\altaffilmark{2},
 Avishay Gal-Yam\altaffilmark{7}, 
  Robert M.~Quimby\altaffilmark{3},
 Shrinivas R.~Kulkarni\altaffilmark{3},
 Ofer Yaron\altaffilmark{7}, and
 David Polishook\altaffilmark{7}
  }

\altaffiltext{1}{Spitzer Science Center/Caltech, Mailcode 220-6,
  Pasadena CA 91125; email: vandyk@ipac.caltech.edu.}
\altaffiltext{2}{Department of Astronomy, University of California,
  Berkeley, CA 94720-3411.}
\altaffiltext{3}{Astronomy Department, California Institute of Technology, Pasadena, CA 91125.}
\altaffiltext{4}{Einstein Fellow.}
\altaffiltext{5}{Institute for Astronomy, University of Hawaii, 2680
Woodlawn Dr., Honolulu, HI 96822.}
\altaffiltext{6}{Hubble Fellow.}
\altaffiltext{7}{Department of Particle Physics and Astrophysics, Weizmann Institute of Science,
  76100 Rehovot, Israel.}

 \begin{abstract}

We have identified a luminous star at the position of supernova (SN) 2011dh/PTF11eon, 
in pre-SN archival, multi-band images of the nearby, nearly face-on galaxy Messier 51 (M51)
obtained by the 
{\sl Hubble Space Telescope\/} with the Advanced Camera for Surveys. This identification
has been confirmed, to the highest available astrometric precision, using a Keck-II 
adaptive-optics image.
The available early-time spectra 
and photometry indicate that the SN is a stripped-envelope, core-collapse Type IIb,
with a more compact progenitor (radius $\sim 10^{11}$ cm) than was the case for 
the well-studied SN IIb 1993J. 
We infer that the extinction to SN 2011dh and its progenitor arises from a low Galactic foreground
contribution, and that the SN environment is of roughly solar metallicity. 
The detected object has absolute magnitude $M_V^0 \approx -7.7$ and
effective temperature $\sim 6000$~K.
The star's radius, $\sim 10^{13}$ cm, is more extended than what has been inferred for the 
SN progenitor.
We speculate that the detected star is either an unrelated star very near the position of
the actual progenitor, or, more likely, the progenitor's companion in a 
mass-transfer binary system. The position of the detected star in a Hertzsprung-Russell
diagram is consistent with an initial mass of 17--19 M$_{\odot}$. The light of this star
could easily conceal, even in the ultraviolet, the presence of a stripped, compact,
very hot ($\sim 10^5$ K), nitrogen-rich Wolf-Rayet star progenitor.
 \end{abstract}


\section{Introduction}\label{intro}

\bibpunct[;]{(}{)}{;}{a}{}{;} 

Determining the stellar origins of supernovae (SNe) is one of the most compelling areas of
modern astrophysics. Progress has been made over the last two decades 
in the identification of the progenitor stars of core-collapse SNe. 
In this {\it Letter}, we consider initial results on the progenitor of SN 2011dh/PTF11eon in M51. SN 2011dh was discovered independently by several amateur
astronomers 
and by the Palomar Transient Factory (PTF) collaboration 
between May 31 and June 1 (UT dates are used throughout), within $\sim$1 day of
explosion; see \citet{arcavi11} and references therein. 
The confirmation spectrum by Silverman et al.~(2011) 
showed a relatively blue continuum and well-developed P-Cygni
profiles in the Balmer series, with 
the H$\alpha$ absorption minimum blueshifted by $\sim 17,600$ km s$^{-1}$, 
indicating the SN was of Type II.
The SN has been further classified as Type IIb \citep{arcavi11}.
SN 2011dh is being intensely studied at a number of wavelengths. 
For instance, the SN was detected early by \citet{margutti2011} with the X-ray
telescope (XRT)
and by \citet{kasliwal11} with the ultraviolet-optical telescope (UVOT) on {\sl Swift\/}; 
by \citet{horesh11a} with the Combined Array for Research in Millimeter-wave 
Astronomy (CARMA); and by \citet{horesh11b} with the Expanded Very Large Array
(EVLA). \citet{soderberg11} also present early panchromatic SN observations. 
The SN position at cm wavelengths is
$\alpha$(J2000) = $13^{\rm h} 30^{\rm m} 05{\fs}104$, 
$\delta$(J2000) = $+47\arcdeg 10\arcmin 10{\farcs}92$ ($\pm 0{\farcs}01$ in each coordinate). 
It is offset $126{\farcs}35$ E and $91{\farcs}70$ S from 
the nucleus of M51a 
\citep[NGC 5194; $\alpha =
13^{\rm h} 29^{\rm m} 52{\fs}711$, $\delta = +47\arcdeg 11\arcmin 42{\farcs}62$;][]{turner94}, 
along a prominent spiral arm.

\citet{chevalier10} have argued that SNe~IIb can arise from progenitor stars with two very different
radii, extended and compact. 
\citet{aldering94} concluded that the SN IIb 1993J progenitor was a massive K0-type supergiant,
with bolometric magnitude $M_{\rm bol} \approx -7.8$.
The radius of the star was extended, $\sim 4 \times 10^{13}$ cm \citep{woosley94}.
The SN 1993J properties could be well explained by an interacting binary system model 
\citep[e.g.,][]{podsiadlowski93,nomoto93,woosley94}; furthermore, the progenitor companion 
appears to have been recovered spectroscopically a decade later \citep{maund04}. 
\citet{arcavi11} report on the detection of the shock-breakout cooling tail for SN 2011dh, which
was also seen for SN 1993J \citep{richmond94}.  
However, the more rapid decline in the 
early light curve and the much lower temperature inferred from the SN 2011dh spectra,
than was observed for SN 1993J \citep[e.g.,][]{filippenko93,swartz93,woosley94},  
indicates that the SN 2011dh progenitor was likely more compact, with radius $< 10^{13}$ cm.
In fact, spectroscopically SN 2011dh more closely resembles the SN IIb 
2008ax \citep{pastorello08,chornock11,taubenberger11}, the progenitor of which 
\citet{chevalier10} estimated to have had a radius $\sim 10^{11}$ cm.
\citet{soderberg11} suggest a radius $\sim 10^{11}$ cm for the SN 2011dh progenitor as well.
We find this evidence for a compact SN 2011dh progenitor to be plausible and have assumed
it to be the case in this {\it Letter}.

\citet{li11a} first identified a progenitor candidate for SN 2011dh in archival 
{\sl Hubble Space Telescope\/} ({\sl HST})
Advanced Camera for Surveys (ACS) images. \citet{li11b} performed
further preliminary studies, measuring initial properties of the candidate. Here
we undertake a more extensive analysis and attempt to constrain the properties 
of the SN 
progenitor system.

The nearly face-on M51, the ``Whirlpool Galaxy,'' 
also hosted the Type I SN 1945A, 
Type Ic SN 1994I, and Type II-Plateau (II-P) SN 2005cs.
We adopt a distance modulus to M51 of $\mu=29.42 \pm 0.27$ mag (distance = 7.66 Mpc), 
from surface brightness fluctuation measurements \citep{tonry01}.\footnote{This 
distance modulus is consistent with $29.41 \pm 0.22$ mag, determined by
\citet{poznanski09} using the SN~II-P standard-candle method 
applied to SN 2005cs, assuming a Hubble constant $H_0 = 70$ km s$^{-1}$ Mpc$^{-1}$.}

\section{Observations and Analysis}

M51 was observed on 2005 January 20--21
by the Hubble Heritage Team (GO/DD 10452; PI: Beckwith) with 
the ACS Wide Field Channel (WFC). A four-band 
(F435W, F555W, F658N, and F814W) image mosaic of M51a
and NGC 5195 (M51b) was obtained in six 
ACS pointings, with four dithered exposures at each pointing. 
These data had been previously analyzed by \citet{li06} to determine
the properties of the SN 2005cs progenitor. The ``drizzled'' mosaics in each band were taken 
from the Hubble Legacy Archive (HLA). 
Individual ``flt'' exposures were also acquired from the {\sl HST\/} Archive.
The approximate location of the SN site was first established comparing the HLA mosaics to 
early-time SN images, obtained with the Katzman Automatic Imaging Telescope 
\citep[KAIT;][]{filippenko01} at Lick Observatory. 

\begin{figure}
\figurenum{1}
\plottwo{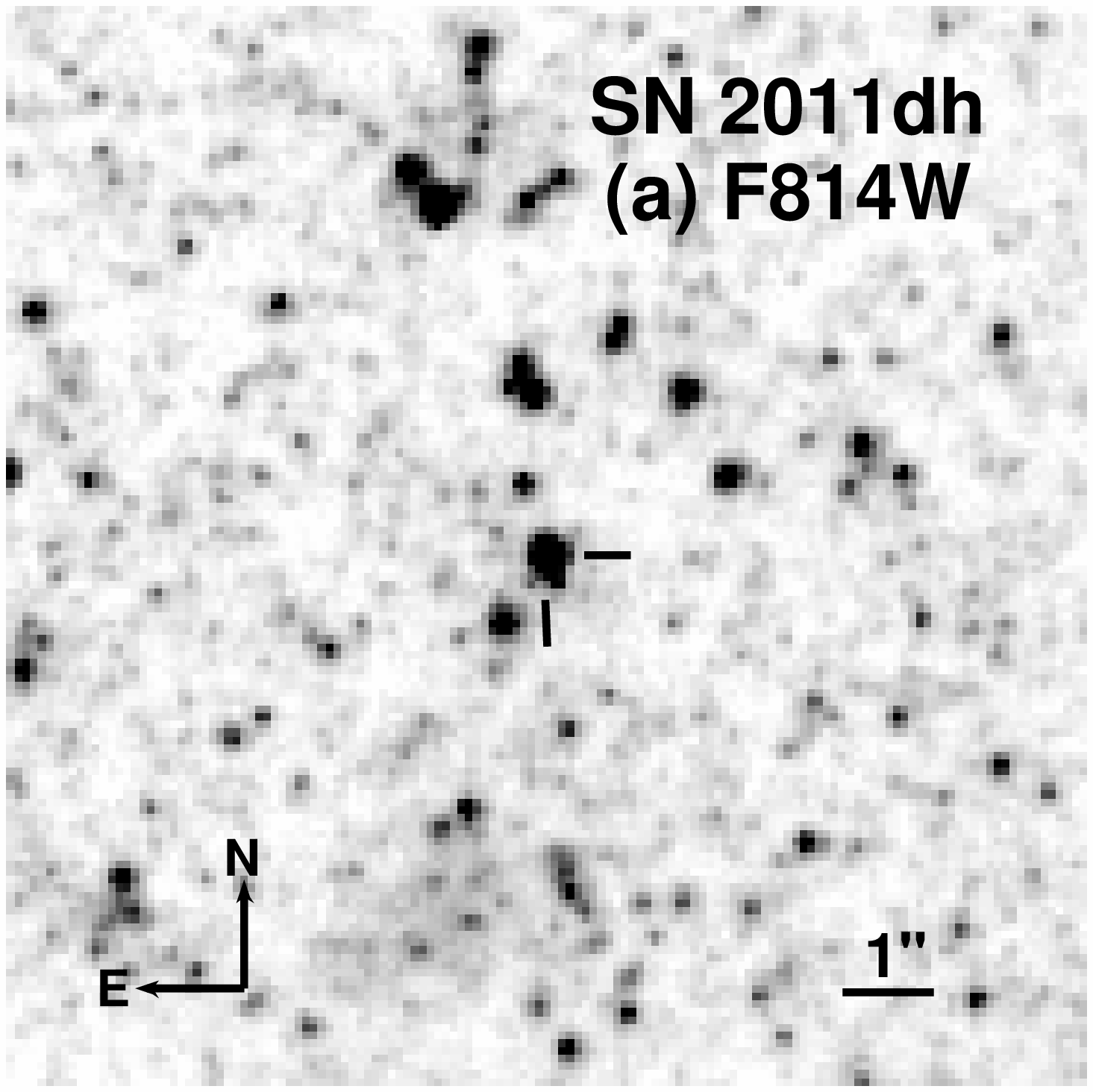}{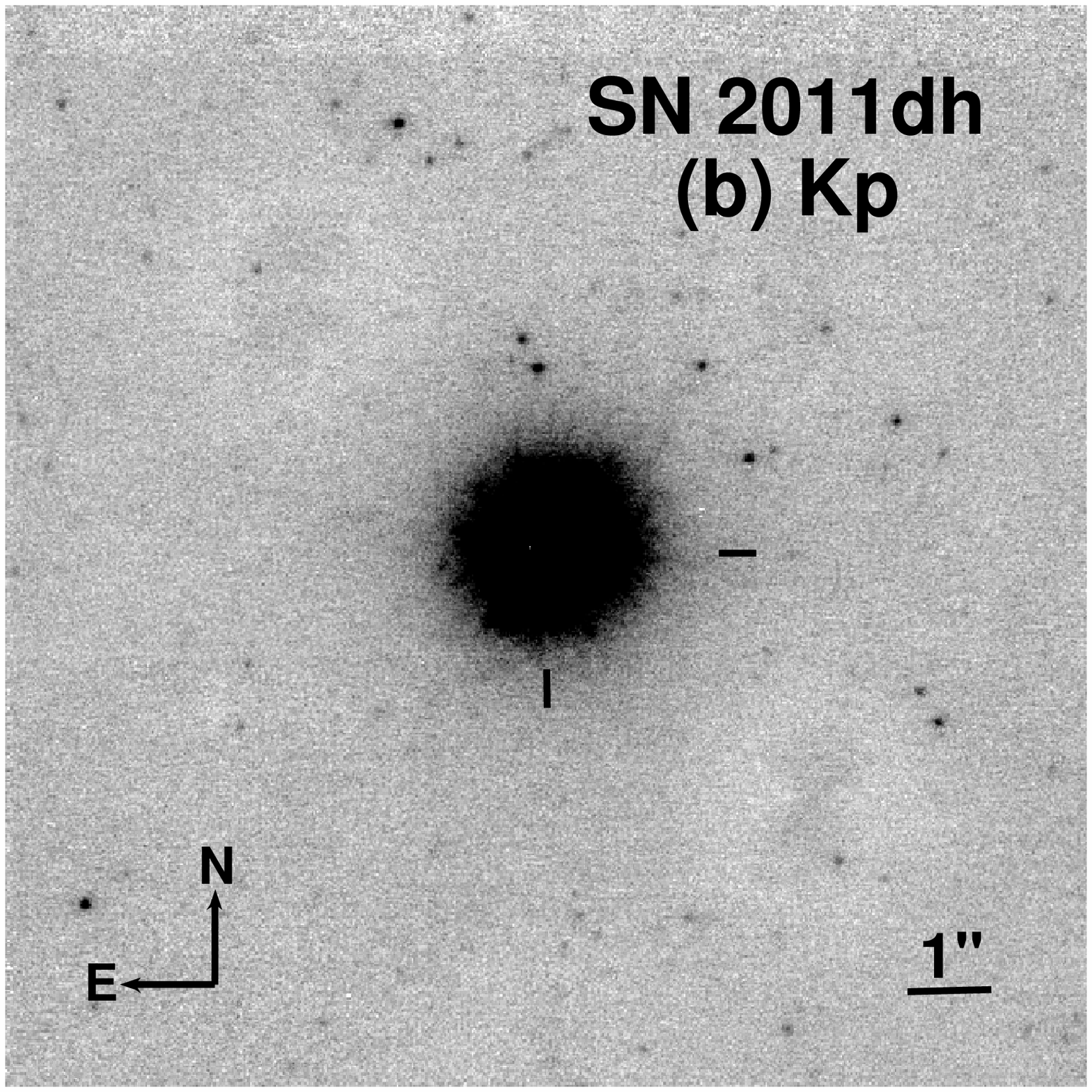}
\caption{(a) A portion of the archival {\sl HST\/} F814W image of M51 from 2005; the
star detected at the precise location of SN 2011dh/PTF11eon is indicated by {\it tickmarks}. 
(b) A portion of the $K_p$-band AO image
obtained using NIRC2 on the Keck-II telescope on 2011 June 6; the SN is indicated by
{\it tickmarks}.\label{figprog}}
\end{figure}

We subsequently (2011 June 6) took
high-resolution, adaptive-optics \citep[AO;][]{wizinowich06} 
$K_p$-band (central wavelength 2.124 $\mu$m; bandwidth 0.351 $\mu$m)
images of the SN, using the Near Infrared Camera 2 (NIRC2) 
instrument on the Keck-II 10-m telescope, to
precisely pinpoint the SN location in the archival {\sl HST\/} data.
The geometric distortion corrections from \citet{yelda10} were first applied to each of the
individual NIRC2 frames before combination of all frames into a single image mosaic.
Measuring the positions using {\it imexamine\/} in IRAF\footnote{IRAF (Image 
Reduction and Analysis Facility) is distributed by the National Optical Astronomy
Observatories, which are operated by the Association of
Universities for Research in Astronomy, Inc., under cooperative agreement
with the National Science Foundation.} of 18 stars seen in common between the 
Keck AO and ACS images, we were able to obtain an astrometric
transformation with formal uncertainty ($\Delta X$, $\Delta Y$) = (0.102, 0.109) pixel, for a total 
root-mean square uncertainty of 0.149 ACS pixel, or 7.45 milliarcsec (mas), 
in the relative astrometry. 
The SN position, seen in Figure~\ref{figprog}b, coincides with the point 
source visible in Figure~\ref{figprog}a. 
Comparison
of the expected ACS pixel position for the SN site, derived from the transformation, to the actual
measured position of the source results in an uncertainty of 0.049 ACS pixel = 2.5 mas,
or $\sim$ 19,000 AU = 0.09 pc (at the distance of M51).
Accordingly, we have very high confidence in the probability 
that this object, the same as the one identified by \citet{li11a}, 
is 
spatially coincident with SN 2011dh.

We analyzed the ACS ``flt'' images in all bands using Dolphot\footnote{The ACS 
module of Dolphot is an adaptation of the photometry package HSTphot (a package 
specifically designed for use with {\sl HST\/} Wide-Field Planetary Camera 2 [WFPC2] 
images; \citealt{dolphin00a}). We used 
v1.1, updated 2010 January 6, from http://purcell.as.arizona.edu/dolphot/, with, e.g., updated ACS
zeropoints.} \citep{dolphin00a}, 
which is especially designed for ACS. We measured the relative offsets 
between the dithered exposures in each band, with respect to one fiducial image, 
before running the package. The output from Dolphot automatically includes the 
transformation from flight-system F435W, F555W, and F814W to the corresponding 
Johnson-Cousins \citep{bessell90} magnitudes in $BVI$, following \citet{sirianni05}.
Since color corrections are required in the \citet{sirianni05} relations to transform
the flight-system broad-band colors into the standard magnitude system, and no such 
color corrections exist for 
F658N, the flight-system magnitude in this narrow 
bandpass could not be transformed to a standard system.
Dolphot indicates with a flag, as well as with measurements of $\chi^2$ and the 
parameter ``sharpness,'' whether a detected source is most likely a {\it bona fide\/} star.
All of these indicators point to the detected object 
being stellar.

The SN site was also imaged in a pair of 1300~s exposures with WFPC2 in F336W on 
2005 November 13 (GO 10501; PI: Chandar). We performed photometry of these images with 
HSTphot. A source is
detected at 4.1$\sigma$ at the SN position in one of the two exposures, but not in the other one. 
The results of all of the photometry for the detected star are in Table~\ref{photprog}.

\section{The Nature of the Detected Star}

\bibpunct[ ]{(}{)}{;}{a}{}{;}

\begin{figure}
\figurenum{2}
\plotone{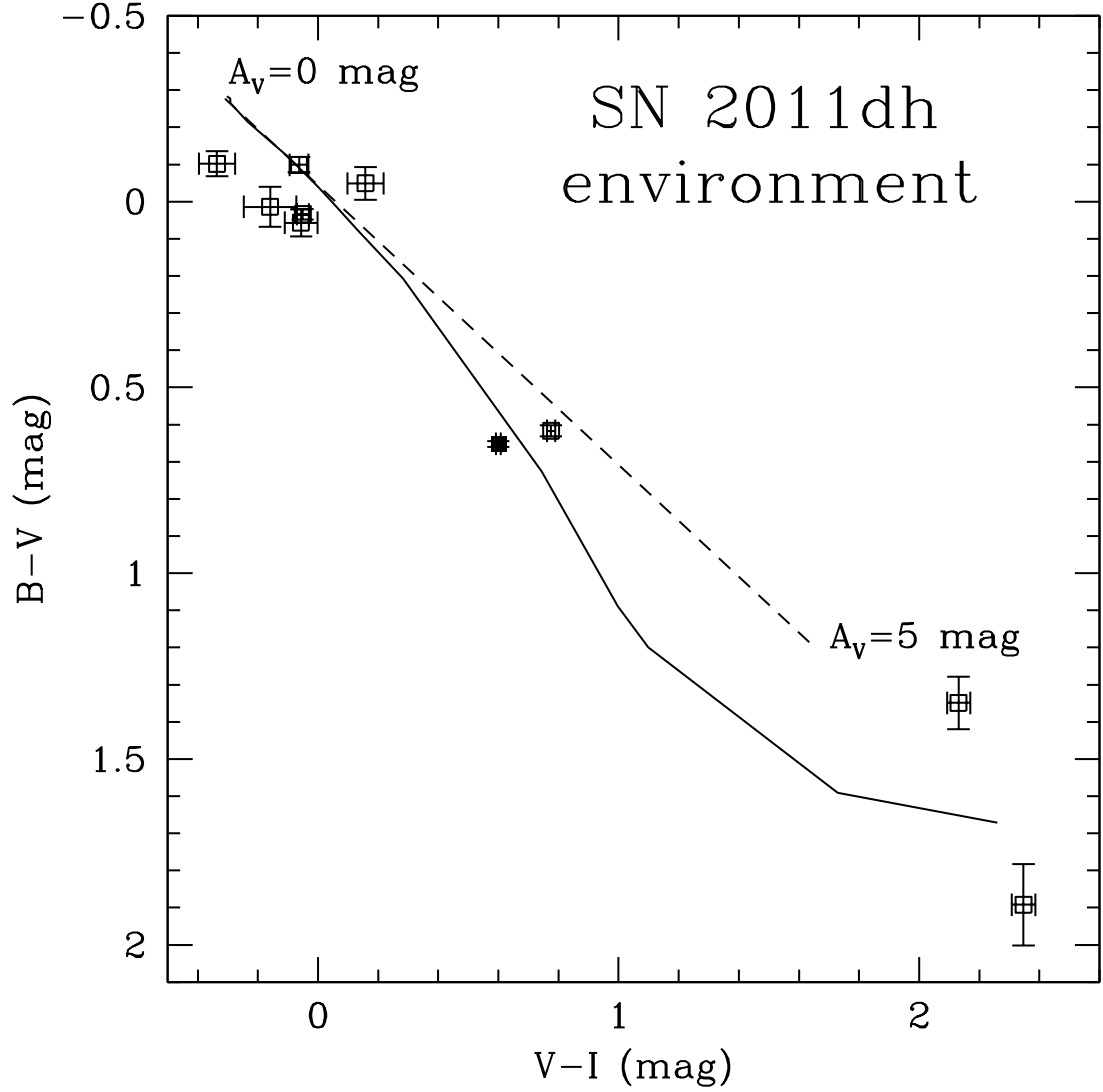}
\caption{Color-color diagram showing the star detected at the precise location
of SN 2011dh/PTF11eon ({\it filled square}),
along with other stars ({\it open squares}) within an area $\sim 91 \times 91$ pc$^2$ 
around this location. Also shown for comparison are the locus of (unreddened) 
supergiants
({\it solid line}), derived from \citet{castelli03} models, and the reddening vector
({\it dashed line}), assuming a \citet{cardelli89} reddening law.\label{figcolor}}
\end{figure}

To estimate the properties of this star, we need 
the M51 distance (\S~\ref{intro}) and the SN extinction. 
To infer the latter, we plot the
star's photometry in a color-color diagram in Figure~\ref{figcolor}. We also show
the colors from Dolphot 
of stars in a $\sim 50 \times 50$ pixel ($\sim 91 \times 91$ pc$^2$) region
centered around the SN site. Furthermore, we show the locus for normal
supergiants 
\citep[synthetic Johnson-Cousins colors extracted 
using the package STSDAS/Synphot within IRAF from 
model supergiants;][]{castelli03}, and the reddening 
vector from the \citet{cardelli89} reddening law.
The detected star, as well as other stars in the immediate
environment, appear to be subject to relatively low reddening. This agrees with the relative lack of
dust emission at the SN site, as seen in pre-SN (2004 May 18 and 22), 
archival {\sl Spitzer Space Telescope\/} images of M51 at 8 $\mu$m (program ID 159; 
PI: Kennicutt). This is also consistent with weak or undetectable
Na\,{\sc i}~D absorption in a SN spectrum obtained on June 3 at the Keck 10-m telescope 
using LRIS 
\citep[however, see][regarding the 
limited utility of this absorption feature in low-resolution spectra to determine SN extinction]{poznanski11}, 
and via a comparison of early SN photometry, obtained using KAIT,
with the dereddened colors of SN 2008ax \citep[e.g.,][]{pastorello08}.
The Galactic contribution to the extinction 
is also relatively low, $A_V=0.12$ mag, 
and $E(B-V)=0.04$ and $E(V-I)=0.05$ mag  \citep{schlegel98}. 
We adopt this foreground value as the extinction toward SN 2011dh.

We also require the metallicity of the SN environment, which 
can be inferred from spectroscopy of H\,{\sc ii} regions 
nearest the SN site.
The regions \citet{bresolin04}
labelled ``53'' (nearest region), ``54,'' and ``55'' have oxygen abundances
12 + log(O/H) = $8.66 \pm 0.09$, $8.49 \pm 0.08$, and $8.60 \pm 0.08$, respectively.
We assume these are representative of the SN environment.
Given that the solar value is 12 + log(O/H) = $8.66 \pm 0.05$ \citep{asplund04} and the average
abundance for H\,{\sc ii} regions in  
the Large Magellanic Cloud is 12 + log(O/H) = $8.37 \pm 0.22$ \citep{russell90}, 
we conclude that the SN 2011dh site is most likely of roughly solar metallicity.

\begin{figure}
\figurenum{3}
\plotone{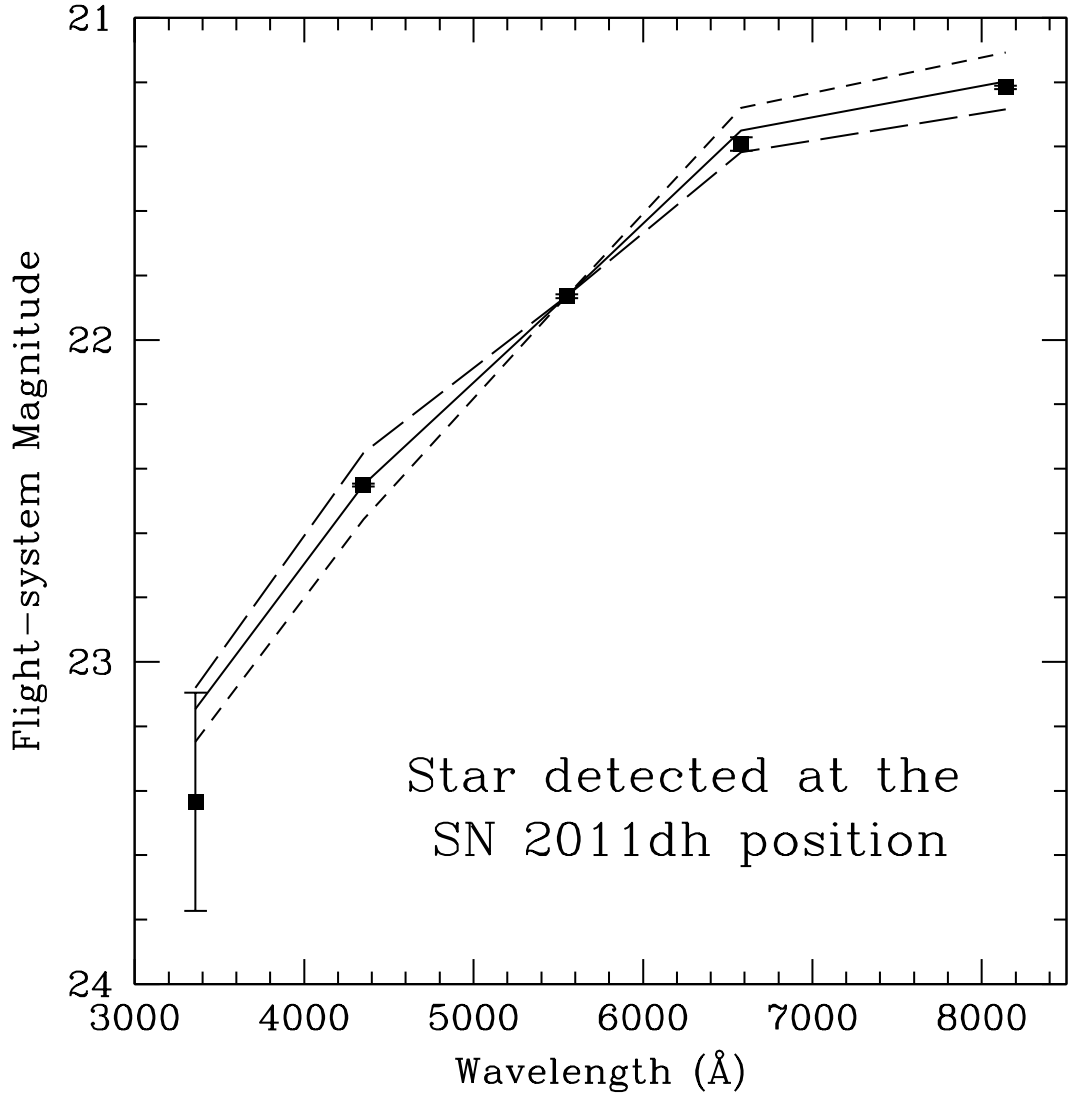}
\caption{Spectral energy distribution (SED) 
of the detected star ({\it points}) in {\sl HST\/} ACS/WFC and WFPC2
flight-system magnitudes. Also shown are the SEDs derived from 
the \citet{castelli03} models for supergiants with effective temperatures
5750 K ({\it short-dashed line}), 
6000 K ({\it solid line}), and 6250 K ({\it long-dashed line}),
after application of the assumed reddening toward the SN 
(see text). The models are normalized to the F555W brightness of the star. \label{figsed}}
\end{figure}

We can estimate the effective temperature, $T_{\rm eff}$, of the detected star 
by modeling its spectral energy distribution (SED) across all observed bands. 
We produced template SEDs via synthetic photometry, extracted using Synphot,
from \citet{castelli03} model stars at solar metallicity with $\log g=1.0$ (see below), 
reddened by our assumed value.
We made the comparison 
in flight-system magnitudes, since the F658N measurement from 
Dolphot and the F336W upper limit from HSTphot 
could not be transformed to a standard system. 
The results are shown in Figure~\ref{figsed}. 
The model with $T_{\rm eff}$ = 6000 K compares well with the observations (particularly
those in the ACS bands). We
therefore adopt this temperature for the observed star, with a conservative uncertainty of
$\pm 100$ K.

From the adopted distance modulus,  
extinction, and reddening, we find that the object
had an absolute intrinsic magnitude of $M^0_V=-7.73$, and intrinsic colors
$(B-V)^0=0.61$ and $(V-I)^0=0.55$ mag. 
We estimate that the probability is $\sim 4 \times 10^{-6}$ that a star more luminous than this
in M51a could be found at this exact location.
From the $T_{\rm eff}=6000$ K model, above, we estimate that the bolometric correction for the 
detected star is $BC_V=0.00$ mag, and therefore $M_{\rm bol}=-7.73$ mag. 
The bolometric luminosity with respect to the Sun 
(assuming $M_{{\rm bol} \odot}=4.74$ mag) is $\log (L_{\rm bol}/{\rm L}_{\odot}) = 4.99$.
The star has a radius of $\sim 290\ {\rm R}_{\odot}$, and we
estimate its surface gravity to be $\log g\approx 0.77$ (assuming the star's mass is 
$\sim 18\ {\rm M}_{\odot}$; see below), so our choice for the model $\log g$ is warranted. 
We note that the star is significantly more extended than a normal supergiant at this temperature
\citep[e.g.,][]{drilling00}.

It is unlikely that the detected star is the one that actually exploded. 
As already noted, the early SN data indicate that the progenitor was far more compact
(radius $\sim 10^{11}$ cm).
Either the detected star is unrelated to the progenitor and is merely a very close
($\lesssim 0.1$ pc) neighbor, or, more likely, it is the companion to the progenitor in a 
binary system. 
The fact that He\,{\sc i} lines were seen in the SN spectra within the first few weeks of explosion 
\citep{marion11} implies that
the progenitor had been substantially stripped of its H envelope, presumably via a wind or
mass exchange with its companion \citep[similar to SN 1993J; e.g.,][]{podsiadlowski93}.

The compact primary star would therefore be very hot, probably a star in a Wolf-Rayet (WR) 
phase.
From Figure~\ref{figsed} one can see that the presumed secondary star alone
can account for much of what was marginally detected in the F336W bandpass.
(Additionally, 
nothing is detected at the SN position in archival {\sl Galaxy Evolution Explorer\/} [{\sl GALEX}] 
ultraviolet NUV and FUV band data for M51.)
A star with characteristics similar to a weak-lined, early-type, N-rich WR, with 
$T_{\rm eff} \approx 10^5$ K, 
$\log (L_{\rm bol}/{\rm L}_{\odot}) \approx 5.3$, and $M_V^0\approx -2.1$ mag 
\citep[adopting the corresponding WNE model from][]{hamann04}, 
would be $\gtrsim 2.2$ mag fainter in the F336W bandpass and would also be 
concealed by its brighter companion in all redder bands. 
This WR star, therefore, would have little effect on the total light of the system, although any WR 
more luminous than this would have been detected at F336W.

\begin{figure}
\figurenum{4}
\plotone{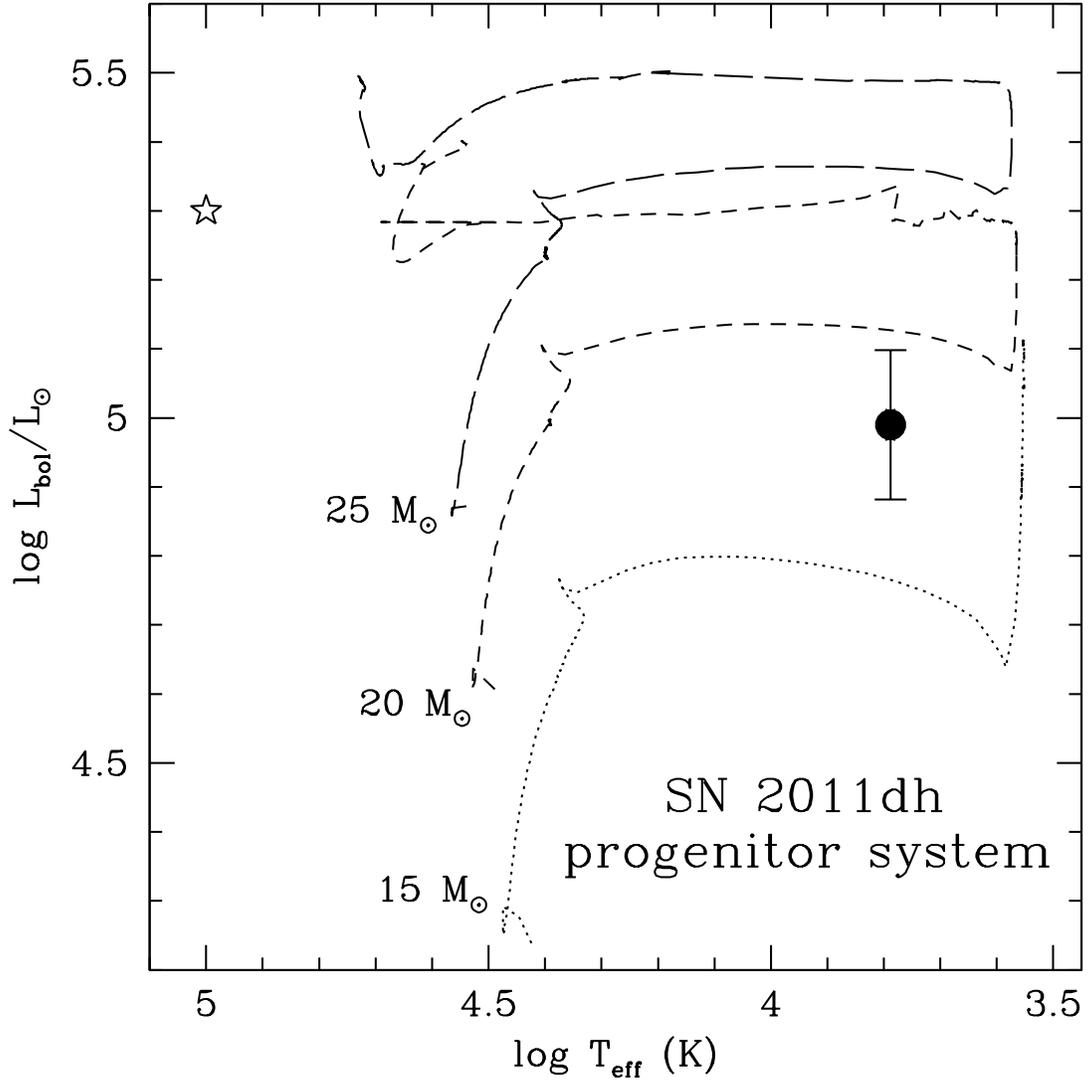}
\caption{Hertzsprung-Russell diagram showing the locus of
  the star detected at the precise location of SN 2011dh/PTF11eon ({\it filled circle}).  
  We speculate that this star is the companion to the progenitor in an interacting binary system.
  We also indicate the locus of a hypothetical compact, hot, Wolf-Rayet primary 
  ({\it five-pointed star}), which we further speculate is the actual SN progenitor.
  Also shown are model single-star evolutionary tracks at
  solar metallicity with equatorial rotation ($v_{\rm rot}=300$ km s$^{-1}$) 
  for initial masses 15, 20, and 25 M$_{\odot}$ \citep{hirschi04}, 
  which are intended merely to be suggestive of the possible mass ranges for
  the two stars.\label{fighrd}}
\end{figure}

In Figure~\ref{fighrd} we show the loci of the detected secondary star and the 
hypothetical primary in a Hertzsprung-Russell diagram.
(The main uncertainty in the luminosity of the detected star arises from the assumed distance 
modulus uncertainty, $\pm 0.27$ mag.)
We also show model stellar evolutionary tracks for massive stars with
equatorial rotation \citep[$v_{\rm rot}=300$ km s$^{-1}$;][]{hirschi04}. These 
single-star tracks are meant to be merely
{\it suggestive\/} of possible masses for the binary components; in particular, 
these tracks do not even adequately account for
the position of the hypothetical primary in the diagram. Clearly, what is required is a full 
modeling of the components of this possible interacting binary system and its evolution up
until the primary's explosion.

\section{Discussion and Conclusions}

\bibpunct[;]{(}{)}{;}{a}{}{;} 

We have detected in archival {\sl HST\/} images a star at the precise location of 
SN 2011dh/PTF11eon.
The star has colors consistent with mid-F-type, although its 
luminosity is higher ($M^0_V\approx -7.7$ mag) 
and its radius more extended ($\sim 290\ {\rm R}_{\odot}$)
than is the case for a normal supergiant. 
The early properties of SN 2011dh, however, point to its having a compact progenitor 
\citep[][ radius ${\sim}10^{11}$ cm, \citeauthor{soderberg11}~\citeyear{soderberg11}]{arcavi11}, 
indicating that the detected star is likely not the star that exploded.
(\citeauthor{maund11}~\citeyear{maund11} favor the interpretation that the star is the yellow 
supergiant progenitor of the SN.)
It is possible that the detected star is just a very close neighbor of, yet generally
unrelated to, the actual progenitor. We consider it more likely, though, that the star is the
companion to the progenitor in a binary system. The extended radius for the detected star and the
compact radius for the progenitor implies that the two stars may have been interacting.
We note, however, that no direct observational indication yet exists that the SN 2011dh 
progenitor was a member of a binary system.

The compact SN 2011dh progenitor and the extended SN 1993J progenitor
are within similar ranges of initial mass \citep[13--22 M$_{\odot}$ for the latter;][]{vandyk02}.
(Note that \citeauthor{crockett08} \citeyear{crockett08} 
concluded that, if the compact progenitor of SN 2008ax were also in a binary system, 
its initial mass range would have been significantly lower, 10--14 M$_{\odot}$.)
We note that many of the 
model SN Ib progenitors experiencing Case A/Case B mass transfer, 
which \citet{yoon10} have recently considered, 
are also in a mass range similar to that suggested
for the SN 2011dh progenitor. 
As both \citet{chevalier10} and \citet{dessart11} point out, the difference between
SN IIb and Ib progenitors could be razor thin, depending on the H mass remaining
in the progenitor star's envelope.

We will, of course, also develop a clearer picture once the SN has significantly faded, 
most likely several years in the future.
At that time, imaging of the SN site can be undertaken, presumably with {\sl HST},
and we can determine if the possible secondary star is still there. For example, 
\citet{ryder06} detected what they concluded to be the late-B- to late-F-type supergiant 
companion to the compact SN IIb 2001ig in ground-based Gemini images
obtained $\sim 1000$ d after explosion.
If the star has vanished or significantly faded, 
we can investigate if any fainter stars
were present, contributing to the observed point-spread function of the star detected in the 
pre-SN {\sl HST\/} images. 
Further work is clearly required to understand more fully
the nature of this interesting, and potentially important, SN.

\acknowledgments 

This work was based in part on observations made with the NASA/ESA
{\it Hubble Space Telescope}, obtained from the Data Archive at the
Space Telescope Science Institute, which is operated by the
Association of Universities for Research in Astronomy (AURA), Inc.,
under NASA contract NAS 05-26555; the W. M. Keck Observatory, which is
operated as a scientific partnership among the California Institute of
Technology, the University of California, and NASA, with generous
financial support from the W. M. Keck Foundation; and the Lick
Observatory, operated by the University of California.  KAIT and its
ongoing research were made possible by donations from Sun
Microsystems, Inc., the Hewlett-Packard Company, AutoScope
Corporation, Lick Observatory, the NSF, the University of California,
the Sylvia \& Jim Katzman Foundation, and the TABASGO Foundation. We
thank the staffs of the Lick and Keck Observatories for their
assistance with the observations.

We thank Peter Nugent for useful comments, and the
referee for helpful suggestions which improved this manuscript. Support for
this research was provided by NASA through grants GO-11575, AR-11248,
and AR-12126 from the Space Telescope Science Institute, which is
operated by AURA, Inc., under NASA contract NAS 5-26555.  A.V.F. and
his group at UC Berkeley also wish to acknowledge generous support from
Gary and Cynthia Bengier, the Richard and Rhoda Goldman Fund,
NASA/{\it Swift} grant NNX10AI21G, NASA/{\it Fermi} grant NNX1OA057G,
NSF grant AST--0908886, and the TABASGO Foundation.  A.G. is supported by the ISF.
E.O.O. is supported by an Einstein Fellowship and NASA grants.  J.M.S. thanks
Marc J. Staley for a graduate fellowship.

{\it Facilities:} \facility{Keck:NIRC2}, \facility{Keck:LRIS},
\facility{HST(ACS)}, \facility{Spitzer(IRAC)}, \facility{GALEX}, \facility{Lick:KAIT}.

\clearpage

\begin{deluxetable}{cccccccc}
\tablewidth{6.8truein}
\tablecolumns{8}
\tablecaption{Photometry of the Star Detected at the SN 2011dh Position\tablenotemark{a}\label{photprog}}
\tablehead{
\colhead{F336W} & 
\colhead{F435W} & \colhead{$B$} & 
\colhead{F555W} & \colhead{$V$} & 
\colhead{F658N} & \colhead{F814W} & \colhead{$I$} \\
\colhead{(mag)} & \colhead{(mag)} & \colhead{(mag)}
& \colhead{(mag)} & \colhead{(mag)}
& \colhead{(mag)} & \colhead{(mag)}
& \colhead{(mag)}}
\startdata
23.434(339) & 22.451(005) & 22.460 & 21.864(006) & 21.808 & 21.392(021) & 21.216(005) & 
21.208\\
\enddata
\tablenotetext{a}{Uncertainties $(1\sigma)$
are given in parentheses as millimagnitudes.}
\end{deluxetable}


\clearpage

\clearpage

\clearpage

\end{document}